# Stochastic Dispatch of Energy Storage in Microgrids: An Augmented Reinforcement Learning Approach

Yuwei Shang, Wenchuan Wu, *Senior Member, IEEE*, Jianbo Guo, Zhao Ma, Wanxing Sheng, Zhe Lv, Chenran Fu

*Abstract*—The dynamic dispatch (DD) of battery energy storage systems (BESSs) in microgrids integrated with volatile energy resources is essentially a multiperiod stochastic optimization problem (MSOP). Because the life span of a BESS is significantly affected by its charging and discharging behaviors, its lifecycle degradation costs should be incorporated into the DD model of BESSs, which makes it non-convex. In general, this MSOP is intractable. To solve this problem, we propose a reinforcement learning (RL) solution augmented with Monte-Carlo tree search (MCTS) and domain knowledge expressed as dispatching rules. In this solution, the Q-learning with function approximation is employed as the basic learning architecture that allows multistep bootstrapping and continuous policy learning. To improve the computation efficiency of randomized multistep simulations, we employed the MCTS to estimate the expected maximum action values. Moreover, we embedded a few dispatching rules in RL as probabilistic logics to reduce infeasible action explorations, which can improve the quality of the data-driven solution. Numerical test results show the proposed algorithm outperforms other baseline RL algorithms in all cases tested.

## NOMENCLATURE

For the management of battery energy storage systems:

| | |
|---|---|
| $SOC$ | Battery state of charge |
| $\sigma$ | battery self-discharge rate |
| $\eta_B$ | battery charging/discharging efficiency |
| $P_B$ | charging/discharging power of the battery |
| $Ca$ | capacity of the battery |
| $V_{B\text{-}nom}$ | rated voltage of the battery |
| $\mathcal{O}$ | lifetime throughput of the battery |
| $N_f$ | number of cycles until failure of the battery |
| $P_B^{\min}, P_B^{\max}$ | minimum, maximum power of the battery |
| $SOC^{\min}, SOC^{\max}$ | minimum, maximum state of charge |
| $c$ | electricity tariff |
| $P_{PCC}^{\min}, P_{PCC}^{\max}$ | minimum, maximum power at PCC |
| $DOD$ | Depth of discharge |

In the Markov decision process:

| | |
|---|---|
| $a, s, R$ | action, state, immediate reward |
| $\mathcal{A}, \mathcal{S}, \mathcal{R}$ | set of all actions, states, rewards |
| Pr | probability transition function |
| $\pi$ | policy (action selection rule) |
| $Q(s,a)$ | value function for taking action $a$ in state $s$ |
| $\alpha$ | step-size parameter |
| $\gamma$ | discount-rate parameter |
| $G_t^{t+n}$ | cumulative rewards from $t$ to $t+n$ |
| $\Phi_{\mathcal{R}}^{a|s}$ | potential function for the domain knowledge |
| $\mathcal{A}_f$ | set of feasible actions |

## I. INTRODUCTION

### A. Background

Volatile energy resources, such as loads from renewable energy based distributed generators (DGs) and electric vehicles (EVs), significantly affect the operation of power systems. In microgrids, we can coordinate volatile energy resources and energy storage to mitigate power fluctuations [1]. Hence, battery energy storage systems (BESSs) are widely used to balance the power and shave peaks in microgrids [2]. Furthermore, BESSs can be scheduled to increase the electricity revenue for microgrid entities by charging energy in low-price periods and discharging energy in high-price periods [3]. Therefore, how to dynamically dispatch the BESS such that the operational costs of the microgrids are minimized while satisfying the operational constraints of the distribution network is a key challenge.

Many studies have focused on the dynamic dispatch of BESSs. Some works employ deterministic optimization models. However, due to the stochastic operation of DGs and EVs, the dynamic dispatch of BESSs is essentially a multiperiod stochastic optimization problem (MSOP). One way to solve MSOPs is to apply scenario-based stochastic programming (SP). In this approach, Monte Carlo simulations are employed to repeatedly generate scenarios across a multistep process. The computational burden increases exponentially with the number of scenarios investigated. Additionally, the life span of a BESS is significantly influenced by its charging and discharging behavior. When incorporating the lifecycle degradation costs of

Manuscript received XX, 2019. This work was supported in part by the National Science Foundation of China under Grant 51725703.

Y. Shang, W. Wu are with Tsinghua University, 100084 Beijing, China (Corresponding Author: Wenchuan Wu, email: wuwench@tsinghua.edu.cn).

J. Guo, Z. Ma, W. Sheng are with China Electric Power Research Institute, 100192 Beijing, China. Z. Lv, C. Fu are with North China Electric Power University, 102206 Beijing, China.

BESSs into the microgrid optimization objectives, considerable cost reduction may be achieved in different applications, e.g., the microgrid planning and operation [4], and the coordinated operation of the BESS and renewable energy [5]. However, most of the existing SP models either assume zero degradation costs for the BESS, or simplify the battery cycle life to a linear function of the Depth of Discharge (DOD), which may introduce additional estimation error on the BESS dispatch cost [6]. When a more accurate degradation cost model is used for BESSs, the MSOP generally becomes nonconvex and computationally challenging [7].

Reinforcement learning (RL) may be a viable alternative for tackling an MSOP with a non-convex objective function [8]. RL arose from dynamical systems theory, and is formalized by the Bellman Equation and Markov decision process (MDP). A fundamental issue in RL is the balance of exploration and exploitation, which facilitates action-value estimation and policy improvement. It is common for a RL agent to occasionally explore some random actions and learn from experience. In Q-learning, this trial-and-error learning process is guaranteed with asymptotic convergence. As the bootstrapping steps increase, the error of action value estimation decreases (i.e., the error reduction property); yet the conventional RL algorithm suffers from increased computation complexity. To reduce the computation burden of the multi-step RL for tackling MSOPs, the Monte-Carlo tree search (MCTS) method may be a viable solution that shows remarkable success recently [9]. Motivated by these achievements, we study how to incorporate MCTS into Q-learning to solve the stochastic dispatch of BESS in microgrids.

### B. Related work

The early related researchers mainly employ deterministic models for scheduling BESSs. Reference [10] introduces linear programing (LP) to mitigate fluctuations in photovoltaic (PV) output and increase the electricity revenue in the microgrid. In addition to increasing the electricity revenue, the efficiency of the BESS is considered in [11], in which a non-linear optimization model is formulated and solved by a meta-heuristic algorithm. Reference [12] formulates a quadratic programming (QP) to achieve economic microgrid dispatch. A different objective is considered in [13], namely to satisfy the constraints of the distribution network by tracking the power profile established on a day-ahead basis. They formulate a QP and employ model predictive control (MPC) to schedule the BESS. These deterministic models neglect the intermittency and variability of volatile energy resources.

Some other researchers formulate the BESS scheduling problem as stochastic optimization models, which tackle the uncertainties associated with volatile energy resources. In [14], a two-stage stochastic mixed-integer programming (SMIP) is formulated to optimize the dispatching policy for microgrids. In [15], the problem of storage co-optimization is addressed by formulating a two-stage SMIP using piecewise-linear approximation of the value function. In [16], the day-ahead scheduling of the BESS in the microgrid is studied. The optimization model incorporates the battery degradation cost using the rainflow algorithm. Yet this work assumes unlimited energy exchange with the distribution network. In [17], a two-stage stochastic mixed-integer nonlinear model is formulated, and the battery degradation cost was considered by simplifying its cycle life as a linear function of the DOD. A similar battery cycle life model is considered in [18]. This work formulates the BESS degradation cost model as an equivalent fuel-run generator, which enables it to be incorporated into a unit commitment problem. In addition to the two-stage SP models, a multistage SP model is formulated in [19], and solved by a customized stochastic dual dynamic programming algorithm.

Besides the above methods, some works have explored RL methods for scheduling BESSs. A deep reinforcement learning method is used in [20] to provide the energy management results for the microgrid. In [21], the Q-learning method is used to optimize the energy management in the microgrid, which considers the variability of stochastic entities. A cooperative RL algorithm is proposed in [7], whose dispatch objectives incorporate a non-convex BESS degradation cost model. In [8], a dual Q-iterative learning algorithm is proposed to minimize the microgrid operation cost. In addition to these studies, RL based solutions are seen in other related problems or fields with promising results. [22] studies a dynamic pricing problem in the microgrid, where the basic Q-learning model is adopted and improved by defining the energy consumption-based approximate state and adopting the virtual experience updates. [23] develops a RL method for the optimal management of the operation and maintenance of power grids. In this solution the tabular Q-learning is used to learn the optimal policy and the neural network then replaces the tabular representation of the state-action value function. However, the RL methods used in these works ignore the uncertainties between state transitions along the multistep bootstrapping trajectories [19].

In the field of machine learning, combining the MCTS method and embedding domain knowledge into data-driven solutions can enhance their performances, which inspire us on tackling MSOPs. For the MCTS algorithms, [24] introduces its basic idea, in which tree search policies are used to asymptotically focus the Monte Carlo trials on multistep bootstrapping trajectories that are promising to high-return rewards. [25] presents a survey of different variants of MCTS. [26] adopts the MCTS to achieve fast multistep simulations in the computationally intensive game GO. In the studies of incorporating domain knowledge, [27, 28] demonstrate the performance enhancement of RL solutions by leveraging different kinds of domain knowledge. To numerically express the rule based domain knowledge, the probabilistic soft logic (PSL) is formalized in [29]. In [30], the PSL is used to map knowledge rules into neural networks. In [31], the PSL is employed to supervise the learning process by knowledge rules.

### Contributions

We formulate a multiperiod stochastic model for dispatching the BESS in microgrids. The degradation cost model of BESSs adopted in this work is a benchmark employed in the microgrid simulation tool HOMER [32] and other applications [33, 34]. In our RL based solution, the key for identifying statistically optimal dispatching policies is the estimation of expected maximum action values. This may be achieved by naively computing the optimal value function in the scenario based search trees containing $b^d$ possible sequences of actions, where $b$ is the number of discretized actions per state (tree's breadth) and $d$ is the number of steps (tree's depth). So, its computation complexity increases rapidly as the number of scenarios increases. To reduce the computation burden, we employ the MCTS algorithm that has made prolific achievements in playing Go. However, from the perspective of game theory, the MCTS in Go tackles two-player zero-sum deterministic game [25], yet in our case it tackles single-player stochastic power dispatching. How to integrate the MCTS into the Q-learning for solving the MSOP is a key challenge in this work. Moreover, in order to incorporate the domain knowledge for performance enhancement, a knowledge incorporation scheme is needed to numerically express different knowledge rules and combine them for reducing infeasible action explorations. The novelty and contributions of our work are two-fold:

1) We propose a RL solution incorporated with MCTS to tackle the MSOP. In this solution, the Q-learning with function approximation is employed as the basic learning architecture that allows multistep bootstrapping and continuous policy learning. To alleviate the computation burden of randomized multistep simulations, a MCTS algorithm is developed to efficiently estimate the expected maximum action values in the iterative learning process.

2) We develop a knowledge incorporation scheme to embed the rules into the learning process. In this scheme, the probabilistic soft logic is adopted to map knowledge rules to potential functions. The potential functions are then combined by soft logic operations to confine the state-wise feasible action space and enhance the performance of the learned policy.

As far as we know, this is the first work of incorporating MCTS and domain knowledge into RL methods in power system applications.

The remainder of the paper is organized as follows. The problem is formulated in Section II. Its solution is given in Section III. The results in case studies are reported in Section IV and the paper concludes in Section V.

## II. PROBLEM FORMULATION

Figure 1 presents a simplified configuration of the problem. The microgrid is connected to the distribution network at the point of common coupling (PCC). Components of the microgrid include DG, EV, other loads, and the BESS. The active power of these components is marked with a positive power flow direction in the figure. For notational convenience, we introduce $P_{SUM}$ to represent ($P_{DG}$ - $P_{EV}$ - $P_{OL}$).

Fig. 1. Simplified architecture of a grid-connected microgrid; dashed arrows define the positive direction of power flow.

### A. Constraints

Let $t$ be the time index. The active power constraint imposed by the distribution utility at the PCC is

$$P_{PCC,t}^{\min} \leq P_{PCC,t} = -(P_{SUM,t} + P_{B,t}) \leq P_{PCC,t}^{\max} \qquad (1)$$

The branch power flow model developed in [35] is adopted for the power flow calculation of the microgrid,

$$p_{ij} + p_j^g - r_{ij}\frac{(p_{ij})^2 + (q_{ij})^2}{(v_i)^2} = \sum_{j' \in \mathcal{J}(j)} p_{jj'} + p_j^d \qquad (2)$$

$$q_{ij} + q_j^g - x_{ij}\frac{(p_{ij})^2 + (q_{ij})^2}{(v_i)^2} = \sum_{j' \in \mathcal{J}(j)} q_{jj'} + q_j^d \qquad (3)$$

$$(v_j)^2 = (v_i)^2 - 2(r_{ij}\,p_{ij} + x_{ij}q_{ij}) + [(r_{ij})^2 + (x_{ij})^2]\frac{(p_{ij})^2 + (q_{ij})^2}{(v_i)^2} \qquad (4)$$

where $i, j$ represent nodes of a line in the microgrid. $p_{ij}$ and $q_{ij}$ are active and reactive power delivered through the line. $v_i$ and $v_j$ are voltage magnitudes. $r_{ij}$ and $x_{ij}$ are resistance and reactance of the line. $p_i^g$ and $q_i^g$ are active and reactive power generation at node $i$. $p_j^d$ and $q_j^d$ are active and reactive power demand at node $j$. $\mathcal{J}(j)$ is the set of all child nodes of node $j$.

The operational constraints in microgrid are given by (5)-(8),

$$v^{\min} \leq v_i \leq v^{\max} \qquad (5)$$

$$(p_{ij})^2 + (q_{ij})^2 \leq (AP_{ij}^{\max})^2 \qquad (6)$$

$$P_{DG}^{\min} \leq P_{DG} \leq P_{DG}^{\max} \qquad (7)$$

$$P_B^{\min} \leq P_B \leq P_B^{\max} \qquad (8)$$

where the node voltage amplitude, power flow of lines, power generation of DGs, and power charging/discharging of BESSs are constrained by their thresholds. Variable $AP_{ij}^{\max}$ denotes the maximum apparent power of the line.

The $SOC$ of the BESS is given by

$$SOC_t = SOC_{t-1}(1-\sigma) - \eta_B \frac{P_{B,t}\,T}{Ca_t V_{B,nom}} \qquad (9)$$

In the charging mode, $\eta_B < 1$ and $P_{B,t} < 0$. In the discharging mode, $\eta_B > 1$ and $P_{B,t} > 0$. $T$ is the time interval.

To prevent damages caused by overcharge/overdischarge, the $SOC_t$ is restricted by

$$SOC^{\min} \leq SOC_t \leq SOC^{\max} \qquad (10)$$

The life-cycle throughput of a BESS is related to the number of operation cycles, $SOC$ in individual cycles, etc. [32],

$$\mathcal{O}_t = (N_f \times e^{\kappa \cdot SOC_t}) \times (1 - SOC_t) \cdot \frac{Ca_{nom}V_{B,nom}}{1000\,W\,/\,kW} \qquad (11)$$

where $\kappa$ is an empirical parameter. The level of BESS degradation is measured by $|P_{B,t}T|/2\mathcal{O}_t$ [33].

### B. Objective function

Assume $t$ is the decision time (which means all state variables are known up to $t$). In order to maximize the operational profits of the BESS, we can formulate the following multiperiod stochastic optimization model,

$$z = \max_a R(s_t, a_t) + \gamma \mathbb{E}_{Pc_{t+1}}[\max_a R(\hat{s}_{t+1}, a_{t+1}) + \gamma \mathbb{E}_{Pc_{t+2}|Pc_{t+1}}$$
$$[\max_a R(\hat{s}_{t+2}, a_{t+2}) + \dots + \gamma \mathbb{E}_{Pc_{t+n-1}|Pc_{t+n-2}}[\max_a R(\hat{s}_{t+n-1}, a_{t+n-1})]\dots]] \qquad (12)$$

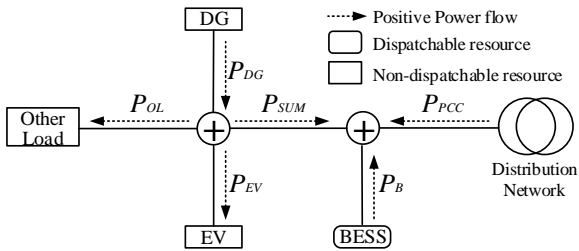

where $R(s_{t+i}, a_{t+i})$ is the immediate reward of taking action $a_{t+i}$ in state $s_{t+i}$. Pr is the probability measure of state transitions and $\mathbb{E}_{\Pr_{t+i}|\Pr_{t+i-1}}$ denotes the expectation taken corresponding to the conditional probability measure ($\Pr_{t+i}|\Pr_{t+i-1}$). $\gamma$ is the discount factor. The superscript $\hat{}$ is used to distinguish stochastic variables from deterministic variables. This notation is also used hereinafter for vectors containing stochastic variables.

In our case, $a_t$ and $s_t$ are given by

$$\begin{cases} a_t = P_{B,t} \\ s_t = (SOC_t, \hat{P}_{SUM,t}, c_t, P_{PCC,t}^{set}) \end{cases} \quad (13)$$

where $c_t$ and $P_{PCC,t}^{set}$ are the time-of-use tariff (TOU) and the active power at $PCC$ expected by the distribution utility, respectively. The BESS is considered as the only dispatchable component. Because $s_{t+i}$ is an unknown future state at decision time $t$, $\hat{P}_{SUM,t+i}$ is modeled as a stochastic variable owing to the volatile energy resources. Because our focus of uncertainty is the power generation/consumption in volatile energy resources, $c_{t+i}$ and $P_{PCC,t+i}^{set}$ are modelled as deterministic variables.

The immediate reward, $R$, contains three factors, defined as

$$R_t = \omega_1 R_{1,t} + \omega_2 R_{2,t} + \omega_3 R_{3,t} \quad (14)$$

where $w_i$ ($i$=1,2,3) are the weights of the different factors. Individual factors are specified as

$$R_{1,t} = \begin{cases} c_{1,t} P_{B,t} T, & P_{B,t} \geq 0 \\ c_{2,t} P_{B,t} T, & P_{B,t} < 0 \end{cases} \quad (15)$$

$$R_{2,t} = -\frac{|P_{B,t}| T}{2 \times Q_{life,t}} C \quad (16)$$

$$R_{3,t} = -|P_{PCC,t} - P_{PCC,t}^{set}| \quad (17)$$

where $R_1$ is the electricity revenue generated by leveraging the TOU tariff. $R_2$ is minus the degradation cost of the BESS due to lifecycle degradation. $C$ is the investment cost of BESS [6, 33]. $R_3$ is minus the penalty cost for power tracking errors at the PCC.

The optimal solution to the above MSOP is a dynamic schedule of multistep charging/discharging actions of BESS. This solution maximizes the microgrid operational benefits indicated by (12)-(17). At the decision time $t$, although only $a_t$ is actually performed, the follow-up scheduled actions can estimate the expected future rewards of $a_t$ more accurately.

## III. Proposed Method

The basic form of RL algorithm is modeled by a tuple $(\mathcal{S}, \mathcal{A}, \mathcal{P}, \mathcal{R})$ in the framework of MDP. $\mathcal{S}$ is the state space containing all state variables. $\mathcal{A}$ is the action space involving all decisions of BESS charging/discharging power. $\mathcal{R}: \mathcal{S} \times \mathcal{A} \in \mathbb{R}$ is the reward function of the state-action pair. $\mathcal{P}: \mathcal{S} \times \mathcal{A} \times \mathcal{S} \rightarrow [0,1]$ is the transition probability from a state-action pair to a successor state, which defines the dynamics of the environment. In deterministic problems, $\mathcal{P} = \boldsymbol{1}$ and $s_{t+1}$ is a deterministic function of state-action pair ($s_t$, $a_t$). However, in stochastic problems such as our case, there is some $\Pr \in \mathcal{P}$ that measures the transition probability. Because predicting the precise transition probabilities of volatile energy is challenging, we develop a RL based approximate solution.

### A. Basic Q-learning architecture with function approximation

To balance the exploration and exploitation in Q-learning, the $\varepsilon$-greedy action selection policy is commonly used,

$$\pi_\varepsilon = \begin{cases} \arg\max_a \mathbb{E}(Q(s_t, a_t)) & \text{with probability 1-}\varepsilon \\ \text{random action} & \text{with probability } \varepsilon \end{cases} \quad (18)$$

where $\varepsilon$ is a small positive value. $\mathbb{E}(Q(s_t, a_t))$ is the expected value of taking action $a_t$ in state $s_t$.

Assume action $a_t$ has been selected in state $s_t$, to estimate its long-term reward we employ $n$-step bootstrapping to update the action value estimates. The cumulative $n$-step future rewards in a bootstrapping trajectory is given by

$$G_t^{t+n}(s_t, a_t) = R_{t+1} + \gamma R_{t+2} + \cdots + \gamma^{n-1} R_{t+n} \quad (19)$$

where $G_t^{t+n}(s_t, a_t)$ is the cumulative action values from $s_t$ to $s_{t+n}$. It is a function of sequential actions $(a_{t+1}, \dots, a_{t+n-1})$ conditioned on $(s_t, a_t)$.

To further incorporate uncertainties for action value updating, we calculate the expectation of $G_t^{t+n}$, i.e. $\mathbb{E}(G_t^{t+n})$. The law for updating the expected action value is

$$\mathbb{E}(Q(s_t, a_t)) \leftarrow \mathbb{E}(Q(s_t, a_t)) + \alpha[\max_{a_{t+1}, \dots, a_{t+n-1}} \mathbb{E}(G_t^{t+n}(s_t, a_t)) - \mathbb{E}(Q(s_t, a_t))] \quad (20)$$

where $\max\limits_{a_{t+1}, \dots, a_{t+n-1}} \mathbb{E}(G_t^{t+n}(s_t, a_t))$ is the expected maximum action value obtained by taking the best-performing actions $(a_{t+1}, \dots, a_{t+n-1})$, following $(s_t, a_t)$.

To allow continuous policy learning, the function approximation is employed in the above tabular Q-learning architecture that achieves a parametric approximation of the action value function,

$$Q(s, a; \boldsymbol{\theta}) \approx \mathbb{E}(Q(s, a)) \quad (21)$$

where $\boldsymbol{\theta} \in \Theta^d$ is a finite-dimensional weight vector. In this work, the basic neural network in [36] is adopted as the function approximator, whose weights can be updated following the gradient descent rule.

The formulation of (20) distinguishes our Q-learning model from [7, 8] that do not incorporate the mechanism of multistep bootstrapping under uncertainty. However, this formulation makes the conventional Q-leaning suffer from increased computation complexity, as more simulation steps and scenarios need to be addressed for estimating the stochastically optimal rewards. We tackle this issue by developing the MCTS algorithm in subsection B.

### B. MCTS algorithm

Different from the MCTS algorithms developed in deterministic games, the MCTS employed in this work needs to incorporate stochastic scenarios into the estimation procedure of expected maximum action values. Here we outline the key ideas of the developed algorithm. More details of this algorithm is explained in the Appendix.

At decision time $t$, the MCTS is applied to estimate $\max\limits_{a_{t+1}, \dots, a_{t+n-1}} \mathbb{E}(G_t^{t+n}(s_t, a_t))$, where the sequential states are represented as tree nodes, and the actions are tree edges connecting different nodes. Let $\{\hat{\omega}_t^{t+n}\}$ be a stochastic vector for the probability distribution of stochastic variables over a planning horizon $n$, we refer to a scenario $\omega_t^{t+n} \equiv \Omega \equiv \Omega_{t+1} \times \cdots \times \Omega_{t+n}$ as a realization (or sampling trajectory) of the stochastic process $\{\tilde{\omega}_t^{t+n}\}$. We then use the notion of $SSP_t$ as a scenario sampling pool for providing the generative scenarios,

$$SSP_t = \{[\hat{P}_{SUM,t+1}^{inf}, \hat{P}_{SUM,t+1}^{sup}]....[\hat{P}_{SUM,t+n}^{inf}, \hat{P}_{SUM,t+n}^{sup}]\} \quad (22)$$

where $\hat{P}_{SUM}^{inf}, \hat{P}_{SUM}^{sup}$ are lower and upper bounds for the confidence interval of $\hat{P}_{SUM}$.

From $SSP_t$, the generative scenarios containing $n$ stochastic variables $\{\hat{P}_{SUM,t+1},....,\hat{P}_{SUM,t+n}\}$ are sequentially sampled that forms different possible scenarios. Let $SSP_t^m = \{\hat{P}_{SUM,t+1}^m,....,\hat{P}_{SUM,t+n}^m\}$ be the $m$th scenario, a search tree is built incrementally that stems from the root node $\hat{s}_{t+1}^m$ and expands from a father nodes $\hat{s}_{t+i}$ to some child node $\hat{s}_{t+i+1}$ ($\hat{s}_{t+1}^m$ is transitioned from $s_t$, at the $m$th scenario). The tree expansion follows the UCT (upper confidence bound for trees) policy

$$\underset{\hat{s}_{t+i+1} \in \text{children of } \hat{s}_{t+i}}{\arg\max} \left( \frac{G_{t+i+1}^{t+n}(\hat{s}_{t+i+1})}{N(\hat{s}_{t+i+1})} + \beta \times \sqrt{\frac{\ln N(\hat{s}_{t+i})}{N(\hat{s}_{t+i+1})}} \right) \quad (23)$$

where $N(\hat{s}_{t+i})$ and $N(\hat{s}_{t+i+1})$ are the visit counts of the father and child nodes, respectively. $\beta$ is a constant variable determining the level of exploration. Initially, (23) prefers nodes with low visit counts. Asymptotically, the nodes that are promising with high values are identified. This policy balances the exploitation of learned value function and the exploration of unvisited nodes.

When a child node is selected in the $m$th scenario, the Monte Carlo rollout policy $\pi_r$ begins at this node and ends at the terminal node $\hat{s}_{t+n}^m$. Each rollout performs a sequential simulation and constitutes $n$ state variables, we use $\{\hat{s}_{t+1}^{m,l},....,\hat{s}_{t+n}^{m,l}\}$ to denote the simulation trajectory in the $l$th rollout. Then the rollout statistics of all traversed edges are backed up,

$$N(\hat{s}_{t+i}^m, a_{t+i}) = \sum_{l=1}^{L} \mathbb{I}(\hat{s}_{t+i}^{m,l}, a_{t+i}) \quad (24)$$

$$Q(\hat{s}_{t+i}^m, a_{t+i}) = \frac{1}{N(\hat{s}_{t+i}^m, a_{t+i})} \sum_{l=1}^{L} \mathbb{I}(\hat{s}_{t+i}^{m,l}, a_{t+i}) \hat{G}_{t+i}^{t+n} \quad (25)$$

where $\mathbb{I}$ is the indicator function. If edge $(\hat{s}_{t+i}^{m,l}, a_{t+i})$ was traversed, $\mathbb{I}(\hat{s}_{t+i}^{m,l}, a_{t+i}) = 1$; Otherwise $\mathbb{I}(\hat{s}_{t+i}^{m,l}, a_{t+i}) = 0$. $\hat{G}_{t+i}^{t+n}$ is the accumulated reward from the node $\hat{s}_{t+i}$ to the end node $\hat{s}_{t+n}$. (24)-(25) updates the visit counts and mean action value function in all simulations passing through that edge.

After $L$ rollouts are executed in the $m$th scenario, we identify the set of best-performing actions and obtain the $n$-step maximum action value for $(s_t, a_t)$,

$$G_t^{t+n}(s_t, a_t \mid SSP_t^m) = R_{t+i}(s_t, a_t, \hat{s}_{t+1}^m) + \gamma \max_{a_{t+1},...,a_{t+n-1}} G_{t+i}^{t+n}(\hat{s}_{t+1}^m, a_{t+1} \mid SSP_t^m) \quad (26)$$

By repeating the above process in different scenarios, the expected maximum action values is approximated as

$$\max \mathbb{E}(G_t^{t+n}(s_t, a_t)) \approx \frac{1}{M} \sum_{m=1}^{M} \max_{a_{t+1},...,a_{t+n-1}} \mathbb{E}(G_t^{t+n}(s_t, a_t \mid SSP_t^m)) \quad (27)$$

where $M$ is the number of scenarios investigated.

There are two differences that distinguish the above MCTS and the MCTS deployed in deterministic games such as Go [25]. The first difference is that in Go only a deterministic scenario is investigated for estimating the value function. In our case, we incorporate different possible scenarios for deriving the expected value function. This is achieved by using the notion of SSP in (22) to allow scenarios generation based on any explicit or implicit probability function, and the expected optimal value are accumulated from individual scenarios by (26)-(27). The second difference is that in Go only the estimated value of the last-stage state (i.e. the terminal node) in each rollout is backed up for updating the value function, which is not an accurate estimation in our case. Thereby, we temporally memorize and accumulate the action values of each transition between father and child nodes by (24)-(25) for updating the value function in each rollout.

### C. Scheme for incorporating knowledge rules

Two definitions are given below to leverage dispatching rules for reducing infeasible explorations in the RL algorithm.

**Definition 1.** Let $\mathcal{K} = \{\varpi_y, k_y(a \mid s)\}_{y=1}^{Y}$ be a set of weighted rule sets, where $k_y(a \mid s)$ is the $y$th rule estimating the feasibility of action $a$ conditioned on state $s$, $\varpi_y$ is the weight of $k_y$.

In practice the knowledge rules can be classified as hard rules and soft rules. Here we consider three rules in the rule set (if desired additional rules can also be included),

$$\begin{cases} k_1(a_t \mid s_t, s_{t+1}): soc^{inf} \leq soc_{t+1} \leq soc^{sup} \\ k_2(a_t \mid s_t, s_{t+1}): P_{PCC}^{inf} \leq P_{PCC,t+1} \leq P_{PCC}^{sup} \\ k_3(a_t \mid s_t, s_{t+1}): \left| P_{PCC,t+1} - P_{PCC,t} \right| \leq P_{Threshold,t} \end{cases} \quad (28)$$

where $k_1$ and $k_2$ are hard rules that require $SOC$ and $P_{PCC}$ to remain within allowable ranges when taking action $a_t$ in state $s_t$ and transitioned to a successor state $s_{t+1}$. The hard rules are definitely not violated, otherwise the security of the power distribution network or the BESS will be damaged. $k_3$ is a soft rule that expects the actual $P_{PCC}$ to have small fluctuations between successive states when taking an action. How to use a soft rule depends on actual needs. For example, when the BESS is funded by an end user who focuses only on electricity revenue, $k_3$ can be relaxed because otherwise some candidate actions with higher rewards will be penalized.

**Definition 2.** Let $\phi_{k_y}(a \mid s)$ be an individual potential function of action $a$ conditioned on state $s$ and examined by rule $k_y$. Let $\Phi_{\mathcal{K}}(a \mid s)$ be the total potential function of action $a$ conditioned on $s$ and examined by the rule set $\mathcal{K}$. Also, let $\mathcal{A}_f$ be the set of feasible action spaces evaluated by $\Phi_{\mathcal{K}}(a \mid s)$.

$\phi_{k_y}(a \mid s)$ can be seen as the numerical expression of rule $k_y$. However, when there exists multiple rules, the logic inferences among them are needed for deriving a final result of the feasibility of candidate actions, especially when these rules are not consistent in evaluating the feasibility of an action. Therefore, we introduce PSL to map knowledge rules into the scalar values taken in the interval [0, 1]. The mapping of $k_y$ into an individual potential function is typically of the form $\phi_{k_y} = (\max\{0, d_{k_y}\})$, where $d_{k_y}$ is a measure of the distance to satisfiability of $k_y$ [29]-[30]. For hard rules $k_y$ ($l$=1, 2), $d_{k_y} = 1$ when the candidate action is evaluated as feasible according to $k_y$, otherwise $d_{k_y} = 0$. For the soft rule $k_3$, an exponential operator is used to measure its distance to satisfiability, i.e.,

$$d_{k_3} = \exp(-\frac{|P_{PCC,t+1} - P_{PCC,t}|}{P_{Threshold,t}}).$$

We then derive the total potential function $\Phi_{\mathcal{K}}(a \mid s)$ from all individual potential functions using certain logic operators. Because we have soft rules that take truth values in [0, 1], the classic Boolean logic is replaced by the Lukasiewicz logic that allows continuous truth values taken from the interval [0, 1]. The logic operators such as AND ($\wedge$), OR ($\vee$), NOT ($\neg$) are redefined as [29]-[30]

$$\begin{cases} \phi_{k_x} \wedge \phi_{k_y} = \max\{\phi_{k_x} + \phi_{k_y} - 1, 0\} \\ \phi_{k_x} \wedge \phi_{k_y} = \max\{\phi_{k_x} + \phi_{k_y} - 1, 0\} \\ \neg \phi_{k_x} = 1 - \phi_{k_x} \end{cases} \quad (29)$$

This redefinition allows a simple and flexible inference among different rules. In this work, let $\bar{\phi}$ be the total potential function of all hard rules, and $\phi$ be the potential function of all soft rules, we have $\bar{\phi} = \phi_{k_1} \wedge \phi_{k_2}$, $\phi = \phi_{k_3}$, and $\Phi_{\mathcal{K}} = \bar{\phi} \wedge \phi$. Hence $\mathcal{A}_f$ is decided by

$$\mathcal{A}_f = \{a \in \mathcal{A} \mid \Phi_{\mathcal{K}}(a \mid s) \geq \sigma_{\mathcal{K}}\} \quad (30)$$

where $\sigma_{\mathcal{K}}$ is the threshold.

### D. The developed RL algorithm

Fig. 2 displays the episodic learning implementations of our RL algorithm.

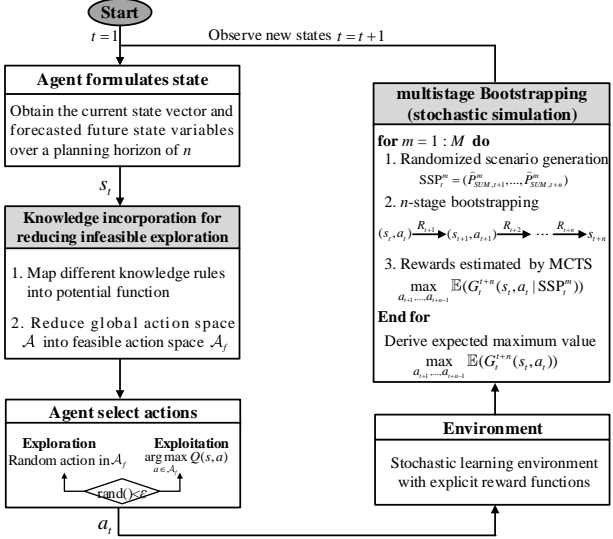

Fig .2 Flow chart of the proposed RL algorithm; two gray boxes highlight the novelty of this algorithm.

At decision time $t$, the RL agent observes its state vector and $n$ stochastic variables $\{\tilde{P}_{SUM,t+1}, ..., \tilde{P}_{SUM,t+n}\}$. Conditioned on these variables, the incorporated knowledge rules are then mapped into the potential function for confining the global action space $\mathcal{A}$ into feasible action space $\mathcal{A}_f$. Then, based on the basic Q-learning framework, the agent selects either an exploitative action $a_t$ with probability 1-$\varepsilon$, or an exploratory action $a_t$ with probability $\varepsilon$ from $\mathcal{A}_{f,t}$. Next, the agent interacts with the stochastic environment and estimates the expected rewards that can be obtained over a $n$-step bootstrapping trajectories. Note that this trajectory starts from the state-action pair $(s_t, a_t)$, and the MCTS is used to sequentially select the remaining $n$-1 actions from $a_{t+1}$ to $a_{t+n-1}$ and estimate the expected maximum cumulative rewards. After simulations of a number of scenarios, the estimated rewards and the parameters of the neural network are updated. The RL agent then continues its learning from the current decision time $t$ towards the next decision time $t+1$, and the above computation process are repeated.

## IV. CASE STUDY

In this section, two microgrid systems are provided to conduct case studies. In Subsection A, a microgrid in [33] is used to verify in detail the performance of the proposed algorithm. In Subsection B, a real microgrid system in China is used to show the effectiveness of the method.

### A. Test case 1

Figure 3 presents the modified microgrid system from [33], which contains two PV systems, two EV charging stations, one BESS, and other loads connected to each node. The rated capacity of the two PV systems, i.e. PV1 and PV2, are 40kW and 20kW, respectively. Two EV charging stations, i.e. EVCS1 and EVCS2, contain 5 AC charging posts and 10 AC charging posts, respectively. The rated power of each charging post is 7 kW. Typical charging modes of EVs include constant current charging, constant voltage charging, etc. The BESS is a 500 kWh lead-acid battery pack. Figure 4 depicts the hourly active power of different components in the microgrid, which shows the high volatility of DGs and EVs. In the stochastic scenarios, the 95% confidence level of $\tilde{P}_{SUM}$ is assumed. For simplicity, $P_{PCC}^{set}$ is set as 50 kW, and the TOU tariffs are referenced from the actual tariffs in China. For the thresholds of the knowledge rules, we restrict the $SOC$ in rule $k_1$ to be within [30%, 90%], $P_{PCC}$ in rule $k_2$ is set to [0, 100 kW], and the variation between the $P_{PCC}$ of two consecutive states in rule $k_3$ is maintained below 50 kW. The training and testing procedures of our algorithm follow [7], [23]. The parameters $\varepsilon$ in (18) and $\beta$ in (22) are set to 1% and 0.7, respectively. The bootstrapping stage is set to 4.

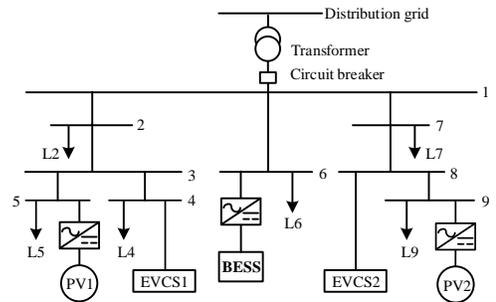

Fig 3 Tested microgrid system; it contains two PV systems, two EV charging stations, one BESS, and 6 residential load points.

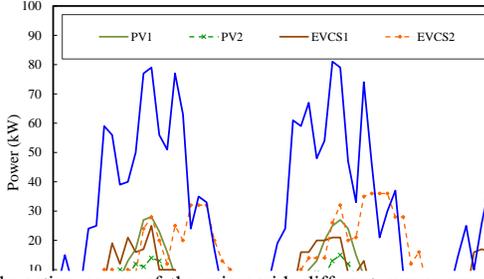

Fig. 4 Hourly active power of the microgrid; different curves are shown in different colors.

We first test the feasibility of the proposed algorithm in realizing its objectives expressed in (12). Figure 5 shows the power management results of BESS in nearly three consecutive days. In sub-figure (a), $P_{SUM}$ fluctuats significantly because of the volatile resources DGs and EVs. In contrast, the dispatching of BESS regulates $P_{PCC}$ for a close tracking of $P_{PCC}^{set}$. Sub-figure (b) shows that the dispatching solution of BESS in general procures energy during low-price low-load periods and sells energy during high-price high-load periods, which increase the electricity revenues. Moreover, a regular charging/discharging behavior of BESS is showed by the $SOC$ curve, thus preventing the accelerated degradation rate caused by over-charging or over-discharging.

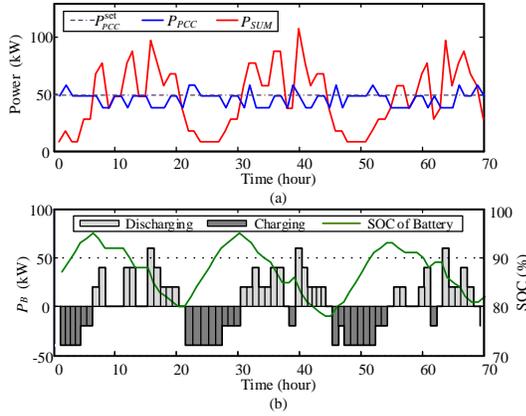

Fig. 5 Power management results of the proposed algorithm. (a) shows the power regulation result at PCC; (b) shows the charging/discharging behavior and the corresponding SOC of the BESS.

We then analyze the computation performance of the developed MCTS algorithm, whose role is mainly to give efficient estimations of the maximum action values over multistep bootstrapping trajectories. To evaluate the degree of accomplishment of this role, we compare the numerical results of MCTS and three algorithms by varying the number of iterations while fixing the investigated scenarios. As listed in Table 1, the compared algorithms include a random search algorithm (RS) that used a random policy during bootstrapping, an exhaustive search algorithm (ES) enumerating candidate actions, and a heuristic search algorithm based on the genetic algorithm (GA). The number of iterations in the numerical tests is varied from $10^1$ to $10^4$. In each iteration budget, we repeat the computations of these algorithms for 10 times and record the mean and variance of different algorithms. The mean values are normalized by the min-max normalization. The variances of BS are omitted because they are zero.

Table 1
Performances of different algorithms in estimating the maximum action values.

| Number of iterations | Mean of maximum action value | | | | Variance of estimation | | |
|---|---|---|---|---|---|---|---|
| | MCTS | RS | GA | ES | MCTS | RS | GA |
| $10^1$ | 0.81 | 0.34 | 0.17 | 0 | 0.62 | 0.99 | 0.49 |
| $10^2$ | 0.92 | 0.42 | 0.26 | 0.18 | 0.34 | 1.56 | 0.11 |
| $10^3$ | 0.97 | 0.38 | 0.34 | 0.42 | 0.10 | 2.00 | 0.01 |
| $10^4$ | 1 | 0.39 | 0.38 | 0.74 | 0.02 | 2.17 | 0 |

The mean value indicates that MCTS is the most efficient algorithm in discovering the maximum multi-step action values. The variance of MCTS is asymptotically reduced as the number of iterations increase, which justifies the robustness and asymptotic convergence of this algorithm. However, ES is the least efficient in estimating the action values. RS is highly stochastic without convergence guarantees regardless of the increase of iterations. For GA, although its variance is the smallest and reached almost 0 after $10^4$ iterations, its estimations of the maximum action value improves slowly when the computation effort increases. One possible explanation is that the iterative searching in GA is stuck in some local optimum after $10^4$ iterations. From above comparisons, we can conclude that the MCTS is the best-performing algorithm for achieving the computation task (20).

We further demonstrate the performance of incorporating the knowledge rules into supervising the Q-learning process. In this test, we compare the proposed algorithm (knowledge incorporation, 4-stage bootstrapping) with two other algorithms, namely Algorithm 1 (no knowledge incorporation, 4-stage bootstrapping) and Algorithm 2 (knowledge incorporation, 1-stage bootstrapping) in terms of the rewards obtained and the actual dispatching results. To increase the learning efficiency when extending the bootstrapping depth from 1 to 4, in our algorithm the immediate reward of one-step state transition is set as the initial value for the follow-up action value updating. Figure 6 depicts the accumulated rewards obtained by these algorithms along the learning trajectory. It shows our algorithm is the most effective one in maximizing the cumulative rewards. Specifically, the advantages in the estimating rewards of our algorithm over Algorithm 1 and 2 are highlighted in the earlier and later learning trajectories, respectively. This result shows that knowledge incorporation is useful when the agent has insufficient experiences. Moreover, extending the bootstrapping depth in conjunction with knowledge incorporation can facilitate the agent to increase its rewards in the long run.

Figures 7 and 8 present further comparison results regarding the actual dispatching performance of these algorithms. In Figure 7, the explorative policy of Algorithm 1 is poor that exacerbate the fluctuations in $P_{PCC}$, but our explorative policy always provide feasible policies that reduce power fluctuations. In Figure 8, Algorithm 2 overdraws the BESS capacity at 67 h, which forces the BESS to charge energy afterwards. Consequently, the power tracking result at the PCC is worsen thereafter. In contrast, our algorithm appropriately manages the SOC and always maintains the power tracking at the PCC. Table 2 compared the actual power management results obtained using the proposed algorithm and Algorithm 2. The results are calculated based on the 72-hour power management results presented in Fig.8. Specifically, the electricity revenue is calculated by equation (15), the BESS degradation cost is calculated by equation (16), and the Standard deviation of $P_{PCC}$ is calculated by $\sqrt{\frac{1}{72}\sum_{t=0}^{71}(P_{PCC,t}-P_{PCC,t}^{set})^2}$, which measures the level of power fluctuation at PCC. For simplicity, the results are shown in per unit values and the results of Algorithm 2 are used as base values. Obviously, our algorithm achieves power tracking with smaller power fluctuations (evidenced by the standard deviation of $P_{PCC}$), and the microgrid gains more tariff revenue with lower BESS degradation costs.

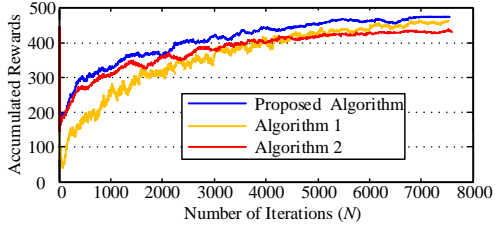

Fig. 6 Comparison of accumulated rewards obtained by three algorithms. The higher the reward, the better the algorithm performance.

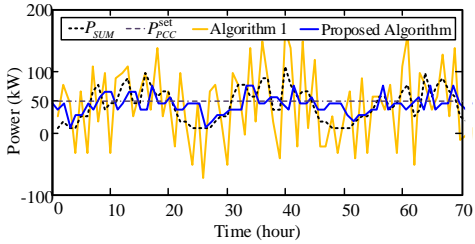

Fig. 7 $P_{PCC}$ regulation results using the explorative policies generated by the proposed algorithm and Algorithm 1.

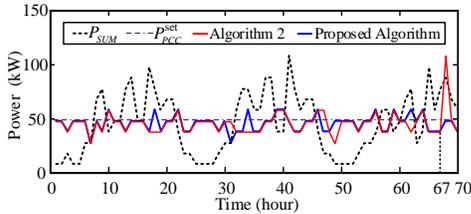

Fig. 8 $P_{PCC}$ regulation results obtained using the optimal policies of the proposed algorithm and Algorithm 2.

Table 2
Dispatch results comparisons of the proposed algorithm and algorithm 2.

|  | Proposed algorithm | Algorithm 2 |
|---|---|---|
| Electricity revenue | 1.05 | 1 |
| BESS degradation cost | 0.92 | 1 |
| Standard deviation of $P_{PCC}$ | 0.79 | 1 |

### B. Test case 2

This test is referenced from a real grid-connected microgrid system installed in Zhejiang, China. Figure 9 shows the configuration and parameters of the test system. It is a hybrid AC/DC microgrid connected to the medium-voltage distribution grid. The AC bus of the microgrid contains 200 kW solar power, 300 kW residential/commercial load, and 500 kW×2 h lead-carbon BESS. The AC bus links the DC bus via a power electronic transformer. The DC bus contains 250 kW solar power,10 kW wind power,250 kW residential/commercial load and 60 kW×2 EV fast charging facilities. We then train and compare the proposed RL algorithm and the baseline RL algorithm (i.e. 1-stage bootstrapping) for dispatching the BESS based on realistic historical load profiles. The aim of the RL agent is to increase the net operation revenue of the microgrid (i.e. the TOU revenue minus the degradation cost of the BESS) while reducing the power fluctuations at the PCC (i.e. measured by $\sqrt{\frac{1}{24}\sum_{t=0}^{23}(P_{PCC,t}-P_{PCC,t}^{set})^2}$). The investment cost of the BESS is ¥ 2/Wh. The TOU tariffs are referenced from the actual tariffs in Zhejiang Province (i.e. ¥ 1.02/kWh from 8:00-22:00; ¥ 0.51/kWh for the rest of the day). Other parameters remain the same as in the test case 1.

Table 3 lists the dispatching results of the two algorithms based on the daily power profiles. The net operation revenue and the standard deviation of $P_{PCC}$ in four days are given. These four days represent different renewable power generation and load consumption patterns in the spring, summer, autumn, and winter, respectively. As can be seen, the proposed method obtains higher revenue with lower power fluctuation at the PCC in all seasons. The biggest gap in revenues is in the autumn, i.e. our method gains ¥ 300.3 more than the baseline method. The largest gap regarding the power fluctuation at the PCC is in the summer, i.e. our method achieves 3% less power fluctuation at the PCC than the baseline method. On average, the daily revenues of our method and the baseline method are ¥ 710.3 and ¥ 560.9, respectively. In the long run, using the proposed method can considerably shorten the cost recovery period for the BESS investor. The above tests provide a first necessary step to prove the effectiveness of the proposed algorithm. Future research efforts will be devoted to test the proposed method on additional numerical models of microgrids.

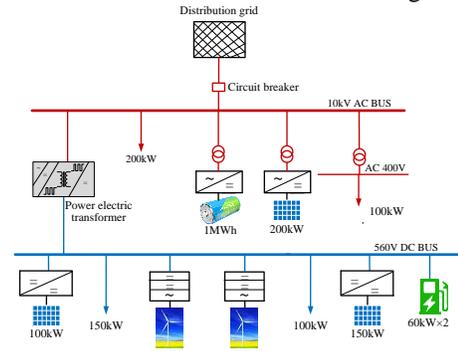

Fig. 9 Configuration of the hybrid AC/DC microgrid system; the parameters of the BESS, distributed energy resources and loads are presented.

Table 3
Result comparisons using based on daily power profiles in different seasons.

| Typical day | | Proposed method | Baseline RL |
|---|---|---|---|
| Spring | Net revenue (¥) | 473.6 | 400.8 |
|  | *SD* of $P_{PCC}$ | 0.161 | 0.169 |
| Autumn | Net revenue (¥) | 873.5 | 543.2 |
|  | *SD* of $P_{PCC}$ | 0.39 | 0.419 |
| Summer | Net revenue (¥) | 1223.7 | 1134.0 |
|  | *SD* of $P_{PCC}$ | 0.138 | 0.168 |
| Winter | Net revenue (¥) | 270.3 | 165.4 |
|  | *SD* of $P_{PCC}$ | 0.242 | 0.266 |

*SD*: standard deviation

## V. Conclusion

In this paper, we present a multiperiod stochastic optimization model for the dynamic management of battery in microgrids. The model is developed to minimize the operational costs of the microgrid, taking into account the nonconvex degradation cost function of the battery energy storage system. Then, we provide a reinforcement learning solution augmented with Monte-Carlo Tree Search and knowledge rules. We first express the knowledge rules into the potential function in the form of soft logic. These knowledge rules are used to confine the state-wise action space, which can reduce the number of infeasible actions explored by the learning agent. To alleviate the computation burden of multistep bootstrapping under uncertainty, the Monte-Carlo Tree search algorithm is modified to increase the estimation efficiency of the expected maximum action values. The results of our numerical tests show that the proposed algorithm asymptotically optimizes the dispatch policy and outperforms other algorithms.

## APPENDIX

The appendix explains how the RL agent learns the dispatching policy in more detail. First, the key steps of the modified MCTS method is explained. Then, the full algorithm of the proposed method is presented.

### A. The MCTS algorithm

To incorporate uncertainties when estimating the cumulative action value for any state action pair, e.g., $(s_t, a_t)$, five steps are needed when performing the MCTS, as shown in Fig. A.1.

**a. Generation.** This step provides randomized sequences containing $n$ sequential stochastic variables, i.e. $[\hat{P}_{SUM,t+1}, \ldots, \hat{P}_{SUM,t+n}]$. The realization of $\hat{P}_{SUM,t+i}$ can be expressed as $\hat{P}_{SUM,t+i} = P^{forecast}_{SUM,t+i} + \Delta\hat{P}_{SUM,t+i}$, where $\Delta\hat{P}_{SUM,t+i}$ is the forecast error. We use the truncated normal distribution (TN) with predefined confidence intervals (CIs) to construct $\hat{P}_{SUM}$ based on the maximum likelihood estimator (MLE), i.e., $\hat{P}_{SUM,t+i} \sim TN(\bar{P}_{SUM,t+i}, \hat{P}_{SUM,t+i})$, where $\bar{P}_{SUM,t+i}$ and $\hat{P}_{SUM,t+i}$ are the sample mean and variance (The details of the TN refer to [9]). Then the Monte Carlo sampling method is used to generate scenarios (the $m$th scenario denotes by $\text{SSP}^m_t$), and (30) is used to form the feasible action space $\mathcal{A}^m_f$.

**b. Selection.** This stage selects explorative policies in the generated scenarios. Given the $m$th scenario, assume the current in-tree simulation step begins at node $\hat{s}^m_{t+1}$ and ends at $\hat{s}^m_{t+n}$, each node $\hat{s}$ of the tree stores the state-action pair $(\hat{s}, a)$, and each edge stores the statistics $\{G(\hat{s}, a), N(\hat{s}, a)\}$, where $N(\hat{s}, a)$ is the visit count and $G(\hat{s}, a)$ is the mean action value for that edge.

**c. Expansion.** This stage incrementally expands the tree until the terminal nodes in a generative scenario. The UCT criterion is used to decide which child node to be expanded. Then the Monte Carlo rollout policy $\pi_r$ begins at this node and ends at a terminal node. During tree expansion, the successively joined leaf nodes result in different combinations of sequential state-action pairs.

**d. Backpropagation.** This stage updates the rollout statistics of each in-tree node backwards from the terminal node to the root node by (24) and (25). After reaching the computation budget (e.g. constraint of iteration, time or memory), the set of state-action pairs with the highest expected rewards is identified as marked in the red rectangle in Fig. A.1.

**e. Update.** This stage updates the action value estimation results for each scenario by (26), and finally accumulate the expected action value estimations of all scenarios by (27).

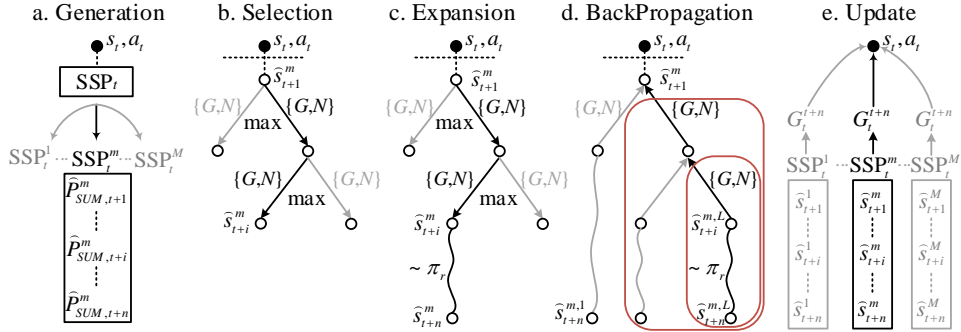

Fig. A.1 Diagram of the modified Monte-Carlo tree search method.

### B. The full algorithm

The proposed approach is shown in Algorithm 1. The main part of this algorithm is shown from line 1 to line 13, where the MCTS method is denoted as the function MCTSSEARCH and realized from line 14 to line 45.

In the main part of the Algorithm 1, given time $t$, the state $s_t$ is observed and the $\varepsilon$-greedy policy is used to select an action $a_t$ from the feasible action space $\mathcal{A}_f$. The $\text{SSP}_t$ is then generated to provide different possible scenarios for the future $n$ time stamps. Given the $m$th scenario, the MCTSSEARCH is performed, whose input parameter $\hat{s}^m_{t+1}$ is transitioned from $(s_t,$

$a_t)$. When $M$ scenarios have been evaluated, the expected maximum action value for $(s_t, a_t)$ can be approximated. This approximated value is marked as the label of a training example, corresponding to input parameters $s_t$, $a_t$, $(\hat{P}^{inf}_{SUM,t+1}, \hat{P}^{sup}_{SUM,t+1}) \ldots (\hat{P}^{inf}_{SUM,t+n}, \hat{P}^{sup}_{SUM,t+n})$. In total, $T$ training examples are provided for learning the weights of the parameterized action-value function (i.e. the neural network) following the gradient descent law.

In the function MCTSSEARCH $(\hat{s}^m_{t+1})$, the subfunctions TREEPOLICY, DEFAULTPOLICY and BACKUP are iteratively executed. In one iteration, the TREEPOLICY determines how to expand the tree from a father node to a child node. In this subfunction, the unvisited nodes are assigned higher priority for node expansion than the visited node selected by the subfunction BESTCHILD. The subfunction DEFAULTPOLICY then performs fast simulations from a current node, e.g. $\hat{s}_{t+i}^{m}$ to the terminal node, and $Q_{\pi_{e}}$ records the cumulative rewards of the simulated trajectory. Afterwards, the subfunction BACKUP updates the cumulative rewards of nodes $\hat{s}_{t+i}^{m},....,\hat{s}_{t}^{m}$ given by the TREEPOLICY and the DEFAULTPOLICY. When the computation budget is reached (e.g., constraint of time, iteration or memory), we identify a complete path of the search tree, with tree edges representing the optimal actions $\{\hat{a}_{t+1}^{m},...,\hat{a}_{t+n-1}^{m}\}$ and tree nodes representing the corresponding states $\{\hat{s}_{t+1}^{m},...,\hat{s}_{t+n}^{m}\}$. Note that in line 31, $f$ denotes the state transition function; in the subfunction DEFAULTPOLICY, the variable $Q_{\pi_{e}}$ is used to sum up the sequential rewards of a simulation trajectory rather than only the terminal reward.

---

### Algorithm 1 Multistep Q-learning incorporated with MCTS

1: Initialize action-value function $Q(s,a;\boldsymbol{\theta})$ arbitrarily
2: **for** $t=1, T$ **do**
3:   observe $s_t$
4:   select $a_t$ from $\mathcal{A}_f$ using the $\varepsilon$-greedy policy
5:   generate $\mathrm{SSP}_t$ according to (22)
6:   **for** $m=1, M$ **do**
7:     sample scenario $\mathrm{SSP}_t^m=\{\hat{P}_{SUM,t+1}^m,....,\hat{P}_{SUM,t+n}^m\}$
8:     perform MCTSSEARCH( $\hat{s}_{t+1}^m$ )
9:     estimate the maximum action value by (26)
10:   **end for**
11:   estimate the expected maximum action value by (27)
12:   update (20) and the weights $\boldsymbol{\theta}$ of the neural network
13: **end for**
14: **function** MCTSSEARCH( $\hat{s}_{t+1}^m$ )
15:   create root node as $\hat{s}_{t+1}^m$
16:   **while** within computational budget **do**
17:     $\hat{s}_{t+i}^m \leftarrow$ TREEPOLICY($\hat{s}_{t+1}^m$)
18:     $Q_{\pi_e} \leftarrow$ DEFAULTPOLICY($\hat{s}_{t+i}^m$)
19:     BACKUP( $\hat{s}_{t+i}^m, Q_{\pi_e}$ )
20:   **return** $a($BESTCHILD$( \hat{s}_{t+1}^m ))$
21: **function** TREEPOLICY($s$)
22:   **while** $s$ is nonterminal **do**
23:     **if** $s$ not fully expanded **then**
24:       **return** EXPAND($s$)
25:     **else** $s \leftarrow$ BESTCHILD($s$)
26:     **return** $s$
27: **function** EXPAND($s$)
28:   choose $a \in$ untried actions from $\mathcal{A}(s)$
29:   add a new child $s'$ to $s$
30:   Initialize $G(s')=0$
31:   $s' \leftarrow f(s,a)$ and $a(s') \leftarrow a$
32:   **return** $s'$
33: **function** BESTCHILD($s$)
34:   **return** $\underset{s' \in \text{ children of } s}{\arg\max} (\frac{G(s')}{N(s')} + \beta \times \sqrt{\frac{\ln N(s)}{N(s')}})$
35: **function** DEFAULTPOLICY($s$)
36:   Initialize $(Q_{\pi_e}, j) = (0,0)$
37:   **while** $s$ is non-terminal **do**
38:     choose random action $a$
39:     $s' \leftarrow f(s,a)$ , $Q_{\pi_e} \leftarrow Q_{\pi_e} + \gamma^j * R(s,a)$ and $j = j+1$
40:   **return** $Q_{\pi_e}$ for state $s$
41: **function** BACKUP($s$, $Q_{\pi_e}$ )
42:   **while** $s$ is not null **do**
43:     $N(s) \leftarrow N(s)+1$
44:     $G(s) \leftarrow G(s) + \gamma * Q_{\pi_e}$
45:     $s \leftarrow$ parent of $s$